\documentclass[sigconf,nonacm]{acmart}

\AtBeginDocument{%
  \providecommand\BibTeX{{%
    \normalfont B\kern-0.5em{\scshape i\kern-0.25em b}\kern-0.8em\TeX}}}

\acmConference[Woodstock '18]{SFDI2019+: Third Workshop on Software Foundations for Data Interoperability}
{October 28, 2019}{Fukuoka, Japan}
\usepackage{listings}
\usepackage{booktabs}
\usepackage{subcaption}

\begin{document}

\title{Analyzing Trade-offs in Reversible Linear and Binary Search Algorithms}
%\title{On Trade-offs in Reversible Search Algorithms}

%
%
%
%
%
%
\author{Hiroki Masuda}
\affiliation{%
  \institution{Nanzan University}
  \streetaddress{Yamazato 18}
  \city{Nagoya}
  \state{Aichi}
  \country{Japan}
}

\author{Tetsuo Yokoyama}
\affiliation{%
  \institution{Nanzan University}
  \streetaddress{Yamazato 18}
  \city{Nagoya}
  \state{Aichi}
  \country{Japan}
}
%\email{tyokoyama@acm.org}

\renewcommand{\shortauthors}{Masuda and Yokoyama}

\begin{abstract}
Reversible algorithms are algorithms in which each step represents a partial injective function; they are useful for performance optimization in reversible systems.
In this study, using Janus, a reversible imperative high-level programming language, we have developed reversible linear and binary search algorithms.
We have analyzed
%, for the first time,
the non-trivial space--time trade-offs between them, focusing on the memory usage disregarding original inputs and outputs, the size of the output garbage disregarding the original inputs, and the maximum amount of traversal of the input. The programs in this study can easily be adapted to other reversible programming languages.
Our analysis reveals that the change of the output data and/or the data structure affects the design of efficient reversible algorithms.
For example, the number of input data traversals depends on whether the search has succeeded or failed, while it expectedly never changes in corresponding irreversible linear and binary searches.
Our observations indicate the importance of the selection of data structures and what is regarded as the output with the aim of the reversible algorithm design.
\end{abstract}

% \begin{CCSXML}
%<ccs2012>
%<concept>
%<concept_id>10011007.10011006.10011008</concept_id>
%<concept_desc>Software and its engineering~General programming languages</concept_desc>
%<concept_significance>500</concept_significance>
%</concept>
%<concept>
%<concept_id>10003752.10003809.10010031.10010033</concept_id>
%<concept_desc>Theory of computation~Sorting and searching</concept_desc>
%<concept_significance>500</concept_significance>
%</concept>
%</ccs2012>
%\end{CCSXML}

%\ccsdesc[500]{Software and its engineering~General programming languages}
%\ccsdesc[500]{Theory of computation~Sorting and searching}

%\keywords{Reversible programs, reversible simulation, linear search, binary search, Janus}

%
%
%
%
%
%
%
%
%
%

%
%
%
\maketitle

\section{Introduction}
%
%
%
%

%Reversible algorithms are useful for exploiting the advantage of reversibility in reversible systems~\cite{Mori17a}.
Reversible algorithms are useful for exploiting the advantage of forward and backward determinism in reversible systems~\cite{Mori17a}.
Their applications include performance optimization in parallel discrete-event simulation (PDES)
and quantum circuit synthesis and design, and
they can be developed using general reversible simulations~(e.g.\ \cite{BuTV01}).
The trade-offs in several simulations have been analyzed in\ \cite{BuTV01} and the references therein; however, their time/space overheads were not negligible. Optimization with human insight leads to improved reversible algorithms for reversible simulations in terms of time, space, output garbage, etc., and such optimizations combining some of these measures have been constructed previously (e.g.\ \cite{AxYo15}).
A memory management method developed exclusively for reversible computing has also been studied (e.g.\ \cite{HaMG17}).
Notably, in the original (irreversible) linear search algorithm, the types of output data and data structure do not affect the asymptotic behavior of the number of traversals of the input data or memory usage. In binary search, there is a trade-off between general solutions and efficient, manually-created reversible programs.

\begin{figure}[b]
\centering
\includegraphics[bb=10 690 400 743,width=.35\textwidth]{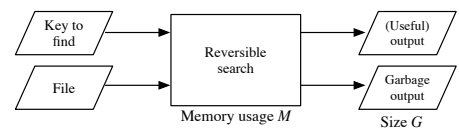}
\caption{
%The input and output of reversible search algorithms to consider.
The I/O of reversible search algorithms to consider.
}\label{fig:inout}
\end{figure}

In this study, efficient reversible linear and binary search algorithms are constructed and the trade-offs between the number of traversals of the input data and the memory usage are analyzed.
As shown in Fig.~\ref{fig:inout}, the reversible search algorithms take a file and a key to find and return
 %one of three types of
 useful outputs such as a flag whether the given record is found, and garbage outputs such as the original input and the control flow information.
They assume different types of outputs and data structures, which do not affect the efficiency of corresponding irreversible algorithms but may do so for reversible algorithms and must therefore be taken into consideration during their design. 
To the best of our knowledge, this is the first attempt to analyze in detail the effect of the selection of the types of outputs and data structures on the efficiency of reversible algorithms. For this reason, the analysis target should be as simple as possible.

To describe reversible programs, Janus, a reversible high-level imperative programming language, was used with procedures and local variable allocations~\cite{YoAG08a}, but the programs presented in this paper can easily be adapted to other reversible languages.
For the design of algorithms high-level languages are more intuitive and expressive than the use of popular computation models such as Turing machines and cellular automata or the low level languages.
The use of a reversible language for reversible algorithm design itself serves as proof that the constructed algorithm is reversible.

\section{Reversible Search}

A search problem is to find a record containing a given key in a collection of a file, i.e.\ $n$ records.
For the sake of simplicity, let each record consist of only a key of a totally ordered domain.
In this study, we only consider as answers of search problems (a) the numbers of records containing the given keys, (b) the locations of the latest records containing the given keys, or (c) flags indicating whether the given keys exists in the given files.

The useful criteria for the optimality of reversible algorithms are expected to be different from irreversible ones~\cite{AxYo15,BuTV01,Fran99}.
We introduce the two criteria for our analysis:
\begin{itemize}
 \item[$M$:] The use of \underline{m}emory, except for the original input and output.
 \item[$G$:] The size of the \underline{g}arbage, except for the original input.
\end{itemize}
Here, the data stored in the memory after the computation, apart from the useful output, is defined as \emph{garbage}.
By this definition, the original inputs are garbage, but in practice, it is not uncommon to use them in further computations.
Therefore, we exclude their count from both $M$ and $G$; most of the programs discussed in this paper retain the files of the original inputs until the end of the computations.
In $G$, the memory for the original input is allowed to be modified during computation if the memory is mutable but should be restored at the end of the computation.
It is possible to realize an irreversible linear search with $M=\Theta(1)$ and $G=0$ because, while it requires some memory to store an index or pointer as well as the output, it can freely erase and overwrite memory at any given time.

%All the reversible search algorithms developed in this study have the same asymptotic time complexity as the corresponding irreversible search algorithms.

%
%
%
%
%

%
%
%
%
%
%
%
%
%

%
%
%
%
%
%
%
%
%
%
%
%
%
%
%

%
%\begin{sloppypar}
%All the programs presented in this paper can be executed on our online interpreter available at \url{http://tetsuo.jp/janus-playground/}.
%\end{sloppypar}

%
\subsection{Reversible Linear Search}
A sequential search from the beginning to the end of the specified files is called \emph{linear}.
The runtime and memory usage of an irreversible linear search is as follows:
The linear search returns a number, location, or flag, as mentioned above, traverses the given input only once, and runs in $\mathrm{O}(n)$ and $\Omega(1)$ time with the memory usage $\Theta(n)$, in which the memory usage except for the original input and output $M$ is $\mathrm{\Theta}(1)$.
In the cases of failure or returning the number of records, the search has to traverse all the records and runs in $\Theta(n)$.
However, in cases where the records are sorted, it can stop after $\mathrm{O}(n)$ and $\Omega(1)$ time.
No garbage is expected $G=0$.

The reversible linear search was manually optimized using domain-specific knowledge and human intuition. Because the optimal solution changes based on the given resource, cases were divided into three categories in terms of the number of input traversals and $G$:
\begin{enumerate}
 \item One traversal and $G=0$,
 \item One traversal and $G \geq 0$, and
 \item One or two traversals and $G = 0$.
\end{enumerate}
(1) is a more efficient special case than (2) and (3), which involve a space--time trade-off.
When $G=0$, the general solution,
saving all the lost information,
%that is otherwise lost
cannot be applied.
When there is only one traversal, the call--copy--uncall scheme cannot be used~\cite{YoAG08a}.

Each file is represented either in (i) an array with a variable containing its size, (ii) a doubly linked list (dlist for short), or (iii) a (singly linked) list.
% and a pointer to the head of the list
% and two pointers to the head and the last of the list.
The locations of the head for (ii) and (iii) and the last for (ii) need to be distinguished from the other records.
Contrary to the irreversible search algorithm, it is seen the output data and the data structure affect the design of the reversible search algorithm.

\textbf{Case (1), (i)--(ii), and (a)--(b).}
Programs were designed with $M=\Theta(1)$.
When an array is used (i), the index is incremented without using extra memory and the index of the last is zero-cleared by the size of the array.
When a dlist
%doubly linked list
is used (ii),
at the end of the computation, the pointer $p$ is zero-cleared by a pointer to the last record $\mathit{last}$. Note that unlike in irreversible setting non-existence of $\mathit{last}$ requires non-constant time overhead to delete the information where the last is located.

The programs must traverse the entire input in case (a), even when the records are sorted beforehand. If the traversal is terminated halfway, the location information must be stored somewhere, but the information cannot be erased without performing additional traversal of the input or adding it to the garbage.

\begin{figure*}[t]
\includegraphics[bb=30 675 1600 745,width=.99\textwidth]{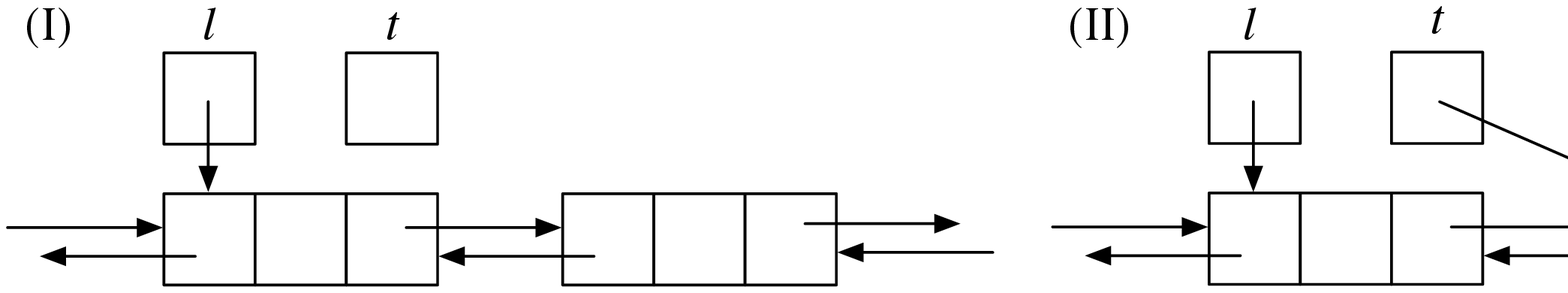}
\caption{Reversible traversal of a dlist.
%doubly linked list.
}\label{fig:pointers}
\end{figure*}

Two example reversible programs are demonstrated.
The program \texttt{srch1} (Case (1), (i), and (b)) in Fig.~\ref{fig:srch1} sets \texttt{i} to the location of the key \texttt{k} (the index of the array) closest to the beginning of the file \texttt{r} of length \texttt{n} if any and \texttt{n} otherwise. Here, that \texttt{n} is returned means that the traversal reaches the test on a sentinel \texttt{r[n]=k} and the search is failed.
\texttt{r[n]} is a sentinel and its value is \texttt{k}.
\begin{figure}[tb]
\begin{lstlisting}
procedure srch1(int r[], int n, int k, int i)
  from i = 0 loop
    i += 1
  until r[i] = k
\end{lstlisting}
\caption{A reversible linear search returning a location using a sentinel.}
\label{fig:srch1}
\end{figure}
The program \texttt{srch2} (Case (1), (ii), and (b)) in Fig.~\ref{fig:srch2} uses a
%doubly linked list
dlist to store a file and returns the location of a record.
\begin{figure}[tb]
\begin{lstlisting}
procedure srch2(int head[], int next[], int prev[], int k, int l)
  from l = 0 loop
    local int t = next[l](*@\label{lst:srch2:allocate}@*)
      l ^= prev[t]^t // update l from prev[t] to t(*@\label{lst:update}@*)
    delocal int t = l(*@\label{lst:srch2:deallocate}@*)
  until next[l]=-1 || k=head[l]
\end{lstlisting}
\caption{A reversible linear search returning a location using a
%doubly linked list
dlist and a sentinel.}
\label{fig:srch2}
\end{figure}
A given key and the result location are stored in \texttt{k} and \texttt{l}, respectively. A
%doubly linked list
dlist is represented by three arrays: 
\texttt{head[]} for keys, \texttt{next[]} for pointers for the next cells, and \texttt{prev[]} for pointers for the previous cells.
For the sake of simplicity, the head record is stored in index 0, and the sentinel is stored at the last of the array. $-1$ indicates a NULL pointer.
In the body of \texttt{srch2}, the location \texttt{l} is repeatedly updated until the last is reached (\texttt{next[l] = -1}) or the key \texttt{k} is found (\texttt{k = head[l]}).
The local clause reserves a memory cell to store an integer value of \texttt{next[l]}, which can be referred by a local variable \texttt{t},
and the delocal clause asserts \texttt{t} has the value of \texttt{l} and frees the memory cell.
The pointer $l$ to a record is updated as shown in Fig.~\ref{fig:pointers}.
%\begin{figure}[t]
%\includegraphics[width=.35\textwidth]{img/pointers.eps}
%\caption{Reversible traversal of a doubly linked list.}\label{fig:pointers}
%\end{figure}
Another pointer $t$ enables the reversible traversal of the records. When we update a pointer $l$ to store an address of a cell to an address of the adjacent cell (I), the address of the adjacent cell is obtained by the data pointed to by $l$, the address is stored in a zero-cleared pointer $t$ at line \ref{lst:srch2:allocate} (II), the address of the original cell is obtained by the data pointed to by the address $t$, and those two addresses are used to update $l$ at line \ref{lst:update} (III), and $t$ is zero-cleared by the same address held in $l$ at line \ref{lst:srch2:deallocate} (IV).
If the dlist is mutable, the list traversal using swap can be performed without another pointer $t$.

\textbf{Case (1), and (iii) or (c).}
Reversible linear search cannot be conducted.
The traversal of
%singly linked
lists requires saving information on which cells have been traversed, and this information cannot be erased in a single traversal.
%
%\paragraph{Case (1) and (c)}
%Reversible linear search cannot be constructed.
A single traversal of the input is not sufficient to erase information on the existence of multiple records having a given key. If such information at the end of computation was a part of garbage, the condition $G=0$ would not be satisfied.

For the case (1), (i)--(ii), and (c), it is possible to construct an algorithm with the further assumption that in an entire given file there is at most a single key that is equal to a given key.
%
%However, it is possible to construct an algorithm in the following cases.
%Note that it is further assumed that in an entire given file there is at most a single key that is equal to a given key.
%Consider a case in which the keys in a file are distinct and consider case (i) or (ii).
It is possible to traverse a given file regardless of the success or failure of the search and set a flag at the time when an answer is found. Therefore, the time $\Theta(n)$ is required for any inputs and $M=\Theta(1)$.

\smallskip
The case (1) has been completely analyzed.  In the following, we only consider the cases which have not been discussed.
%thus far.
%
%
\smallskip

\textbf{Case (2), and (i)--(ii).}
When the search has succeeded and been terminated, the location can be stored and returned as garbage ($M = G = \Theta(1)$).
When the search has failed, the location can be zero-cleared ($M=\Theta(1), G=0$).
No additional space is necessary except to store the garbage.

\textbf{Case (2) and (iii).}
The traversal of immutable lists requires memory space proportional to the number of records traversed.
In the case of locations (b) or flags (c) the search can be terminated halfway ($G=\mathrm{O}(n)$). In the case of the numbers (a), the whole list must be traversed ($G=\Theta(n)$), but if the file is sorted, the search can be terminated halfway ($G=\mathrm{O}(n)$).
No additional space is necessary except to store the garbage ($M=G$).
Nonetheless, the
% singly linked
list carries significant linear space overhead.

If the list is mutable, it is possible to traverse it by replacing pointers and thus create the reversed list in the process.
The original inputs are destroyed, and the remaining list, the reversed list, and location become the garbage output.
This technique is useful when the original input is no longer necessary in the subsequent computation.

\smallskip
\textbf{Case (3), (i)--(ii), and (a).}
The records are traversed entirely regardless of success or failure.
If the given list is sorted, the call--copy--uncall scheme can be performed in $\mathrm{O}(n)$ time.
%used, as in the previous case.

\textbf{Case (3), (i)--(ii), and (b)--(c).}
Here, the call--copy--uncall scheme is only used in the case of success.
When the search succeeds, it requires two traversals of the input and $\mathrm{O}(n)$ time.
When the search fails, the location information is zero-cleared.
The failure always runs in $\Theta(n)$.
The number of traversals differs based on the success or failure.
% of the search.
This is specific to the reversible version of search.

If the given list is sorted and the call--copy--uncall scheme is used in both success and failure cases, it can run in $\mathrm{O}(n)$ time.

\textbf{Case (3) and (iii).}
List traversal requires saving the locations of previous cells (size $\Theta(n)$) and eventually erasing this information through a second traversal of the input.
If the given list is mutable, we can update the pointers of traversed cells and create an intermediate reversed list to rewind the computation ($M=\Theta(1)$).

If the given list is immutable, extra space is required ($M=\Theta(n\log n)$), where $\Theta(\log n)$ is the size of each pointer.
If the given list is sorted, the order notation $\Theta$ can be replaced with $\mathrm{O}$, as in the above cases.

For example, the program \texttt{srch3} in Fig.~\ref{fig:srch3} is a reversible linear search. Formal parameters of the same names as Fig.~\ref{fig:srch2} are used for the same purpose. Variable \texttt{f} is a flag value indicating the success or failure of the search.
\begin{figure}[tb]
\begin{lstlisting}
procedure srch3(int head[], int next[], int prev[], int k, int f)
  local int l = 0
    call srch2(head,next,prev,k,l)
    if l = size(head)-1 then  // search failed
      l ^= size(head)-1       // zero clear l
    else
      f ^= 1                  // search succeeded
      uncall srch2(head,next,prev,k,l)
    fi f != 1
  delocal int l = 0
\end{lstlisting}
\caption{A reversible linear search returning a flag using a
%doubly linked list
 dlist.}
\label{fig:srch3}
\end{figure}
In the body of \texttt{srch3}, \texttt{srch2} is called and the cases are divided by the success and failure of the search.
\texttt{size(c)} returns the length of the array \texttt{a}.
The failure of the search is detected when \texttt{l} points to the sentinel, which is stored at the last of the array as mentioned above.
When the search has failed, \texttt{l} is zero-cleared by the index of the last.
When the search has succeeded, the flag \texttt{f} is set to $1$ and the location \texttt{l} is zero-cleared by the inverse computation of \texttt{srch2}.
The number of traversals of the input differs based on the failure or the success of the search.

\smallskip
The measures $M$ and $G$ are analyzed in all the cases for (1)--(3), (i)--(iii), and (a)--(c), and summarized in Table \ref{tab:comparison}.
The measures change for each data, output, and resource constraint.
These differences do not occur in irreversible corresponding programs.
NA indicates that there is no algorithm for the specified case.

\begin{table}[t]
\caption{Efficiency of reversible linear searches.}\label{tab:comparison}
\vspace{-.3cm}
\begin{minipage}{0.49\textwidth}
\centering
 (1) One traversal and $G=0$.\\
 \begin{tabular}{l|c|c|c}
 \toprule
  $M$ & (a) number & (b) location & (c) flag \\
  \midrule
 (i) array/(ii) dlist & $\Theta(1)$ & $\Theta(1)$ & NA \\
% (ii) doubly-linked list & NA & $\Theta(1)$ & $\Theta(1)$\\
 (iii) list & NA & NA & NA \\
 \bottomrule
 \end{tabular}
\end{minipage}
~
\\[1mm]
\begin{minipage}{0.49\textwidth}
\centering
 (2) One traversal and $G\geq 0$.\\
 \begin{tabular}{l|c|c|c}
 \toprule
 $M, G$ & (a) number & (b) location & (c) flag \\
  \midrule
 (i) array/(ii) dlist & $\Theta(1), \mathrm{O}(1)$ & $\Theta(1), \mathrm{O}(1)$ & $\Theta(1), \mathrm{O}(1)$ \\
% (ii) doubly-linked list & $\Theta(1), \mathrm{O}(1)$ & $\Theta(1), \mathrm{O}(1)$ & $\Theta(1), \mathrm{O}(1)$ \\
 (iii) list & $\Theta(n), \Theta(n)$ & $\mathrm{O}(n), \mathrm{O}(n)$ & $\mathrm{O}(n), \mathrm{O}(n)$ \\
 \bottomrule
 \end{tabular}
\end{minipage}
~
\\[1mm]
\begin{minipage}{0.49\textwidth}
\centering
 (3) One or two traversals and $G=0$.\\
 \begin{tabular}{l|c|c|c}
 \toprule
 $M$ & (a) number & (b) location & (c) flag \\
  \midrule
 (i) array/(ii) dlist & $\Theta(1)$ & $\Theta(1)$ & $\Theta(1)$ \\
% (ii) doubly-linked list & $\Theta(1)$ & $\Theta(1)$ & $\Theta(1)$ \\
 (iii) (mutable) list & $\Theta(1)$ & $\Theta(1)$ & $\Theta(1)$ \\
 (iii) (immutable) list & $\Theta(n)$ & $\mathrm{O}(n)$ & $\Theta(1)$ \\
 \bottomrule
 \end{tabular}
\end{minipage}
\end{table}

\subsection{Reversible Binary Search}
A binary search targets the records sorted according to the key sizes.
It compares the given key and the median record; if they are not equal, the process is repeated within the half of the records that may include the given key.
An efficient binary search runs in $\mathrm{O}(\log n)$ time.

In each step of an irreversible binary search, the information of the current range is not sufficient to reconstruct the previous range, and the output is not sufficient to identify the final range. A general reversible simulation such as classifying this information as garbage makes the program reversible, but it is inefficient in manipulating the additional space of size $\mathrm{O}(\log_2 n)$.

Here, the program is made reversible based on the observation that
if the range is a power of $2$, the previous range is determined up to unique.
This is because the previous range is twice as large as the current one, and at the $j~(\geq 1)$ th iteration when the $\lceil \log_2 n\rceil -j$ th bit from the least significant bit is 0, the former half is selected, or else, the latter half is selected.
Moreover, the search is terminated only if the range size is 1. Hence, the range at the final state is determined up to unique by the output.

In this section, we consider a reversible binary search that takes a file stored in an array, searches for a key and returns the indices of the matching records if the search succeeds or $-1$ if it fails.

A reversible binary search program is shown in Fig.~\ref{fig:bsrch}.
\begin{figure}[t]
\centering
\begin{lstlisting}[texcl=true]
// set the ceiling of logarithm of n to v
procedure log2ceil(int n, int v)
  from v = 0 loop v+=1 until 2**v >= n

procedure bsrch1(int in[], int u, int k, int len)
  local int l = 0
    local int i = len
      u ^= 2**i
      from l=0 && u=2**len loop
        i -= 1
        local int m = 0
          m ^= l + (u-l)/2(*@\label{lst:bsrch1_middle}@*)
          if size(in)<=m || in[m]>k then
            u ^= (m-l)*2 + l       // zero clear u
            u <=> m                // zero clear m
          else
            l ^= 2*m - u           // zero clear l
            l <=> m                // zero clear m
          fi (l&(2**i)) = 0(*@\label{lst:bsrch1_assert}@*)
        delocal int m = 0
      until i = 0
    delocal int i = 0
    u -= 1(*@\label{lst:dec_u}@*)
  delocal int l = u(*@\label{lst:free_l}@*)

procedure bsrch(int in[], int k, int u)
  local int len = 0
    call log2ceil(size(in),len)   // set len\label{lst:call_log2ceil}
    call bsrch1(in,n,u,k,len)
    if u >= size(in) || in[u]!=k then(*@\label{lst:judge}@*)
      uncall bsrch1(size(in),n,u,k,len)
                                  // zero clear u
      u ^= -1
    fi u = -1
    uncall log2ceil(size(in),len) // zero clear len\label{lst:uncall_log2ceil}
  delocal int len = 0
\end{lstlisting}
\caption{A reversible binary search.}\label{fig:bsrch}
\end{figure}
\texttt{in[]} is an array of records which consist only of keys,
\texttt{k} is a key to be found, and
\texttt{u} stores an answer location or dummy output indicating the failure of the search.
Calling \texttt{bsrch(in,k,r)} sets to \texttt{u} the index of key \texttt{k} in \texttt{in[]} if any, and $-1$ otherwise.
Only \texttt{u} changes before and after the call.
Therefore, $G=0$.

We compute $\lceil \log_2 n \rceil$, i.e.\ a smallest integer that is larger than or equal to $\log_2 n$
by means of the call--copy--uncall scheme
where the call \texttt{log2ceil} sets to zero-cleared \texttt{len} at line \ref{lst:call_log2ceil} and 
the uncall is performed to zero-clear \texttt{len} at line \ref{lst:uncall_log2ceil}.

The reversible procedure \texttt{bsrch1} iteratively narrows the searched range in the reversible loop.
The range is represented by the indices \texttt{l} to \texttt{u-1}.
Before the loop, \texttt{l} is set to $0$, \texttt{u} is set to $2^{\lceil\log_2 n\rceil}$.
Thus, the range size is expanded from \texttt{n} to the smallest power of $2$ that is greater than or equal to \texttt{n}.
At line \ref{lst:bsrch1_middle} the middle of the range is set to \texttt{m}.
Next, to halve the range, the reversible conditional decreases the upper index \texttt{u} if the middle index is out of the range of \texttt{in[]} or the middle \texttt{m} is greater than the given key \texttt{k}; otherwise, it increases the lower index \texttt{l}.%
\footnote{Tests and assertions are short-circuit expressions. Therefore, even if the \texttt{m} th of \texttt{in[]} does not exist, \texttt{n <= m} passes the control to the then branch.}
Only the \texttt{else} branch in the $j~(\geq 1)$ th iteration changes the $\lceil\log_2 n\rceil-i$ th bit from the least significant bit from $0$ to $1$. Therefore, the control is merged at line \ref{lst:bsrch1_assert}.
After the loop, the range size is just one (\texttt{l} equals to \texttt{u-1}).
\texttt{u} is set to the index at line \ref{lst:dec_u} and \texttt{l} is deallocated at line \ref{lst:free_l}.

At line \ref{lst:judge}, the procedure \texttt{bsrch} checks whether \texttt{in[u]} is a record to search.
If not, \texttt{u} must not be a part of the output and should be erased.
In such a case, the inverse call of the reversible procedure \texttt{bsrch1} zero-clears \texttt{u}, and
\texttt{u} is set to $-1$, indicating that the search has failed.
Because the index must be non-negative, the assertion \texttt{u = -1} merges the control flow.

When the search fails, \texttt{bsrch1} is called twice (once for call and once for uncall), and the input array is traversed twice. However, a single traversal is sufficient in the case of a successful search.
This serves as an advantage over the general solution using the call--copy--uncall scheme.

\texttt{bsrch} runs in $\Theta(\log n)$ time and $M=\Theta(1)$.
It should be noted that the initial expansion of the search range to a maximum of twice the size does not require the allocation of additional memory cells, and the asymptotic time/space complexity are not degraded.

\section{Conclusion}
%\textbf{【TODO: Elaborate the limitations of the method and implications of the results. What factors were not taken into consideration during this algorithm design? What are the direct and indirect applications of reversible programming taking these trade-offs into consideration? This type of information gives important context to this work.】}
%
%
%

%
%
%
%
%
%
%
%
%
%

In this study, purely reversible implementations of reversible linear and binary searches are presented, and it is shown that the types of output data and data structures used, which do not affect the efficiency of irreversible algorithms, can affect the efficiency of corresponding reversible algorithms.
It is demonstrated here that there are trade-offs in terms of the amount of traversal of the input data, the memory usage $M$, and garbage size $G$ in reversible linear and binary searches of multiple types of outputs and data structures.
Because linear and binary searches are fundamental, a number of the programs in trade-off relationship are useful in reversible software development under various resource conditions.

The design of an efficient reversible algorithm is not trivial; for example, reversible linear (resp.\ binary) search requires two (resp.\ one) traversals in the case of success and one (resp.\ two) traversal in the case of failure.
This difference is not directly implied by the existing design of conventional linear and binary searches.
Clearly, the design of reversible algorithms requires different principles from the design of conventional (irreversible) algorithms.

Our work has targeted linear and binary searches.
In conventional algorithm design, more general strategies such as dynamic programming and greedy algorithms have been proposed.
Developing such general strategies in reversible setting is one of our future work.
The criteria used in this paper $M$ and $G$ are useful for analyzing reversible linear and binary search algorithms.
However, to what reversible algorithms it is useful to analyze is not clear and more meaningful criteria for more general purpose is desirable.
In the future, we aim to further develop the criteria for measuring the efficiency and design techniques of reversible programs.

\begin{acks}
This work was supported by JSPS KAKENHI Grant Number 18K11250 and Pache Research Subsidy I-A-2 for the 2019 academic year.
\end{acks}

%\bibliographystyle{ACM-Reference-Format}
%\bibliography{%

\begin{thebibliography}{6}

%%% ====================================================================
%%% NOTE TO THE USER: you can override these defaults by providing
%%% customized versions of any of these macros before the \bibliography
%%% command.  Each of them MUST provide its own final punctuation,
%%% except for \shownote{}, \showDOI{}, and \showURL{}.  The latter two
%%% do not use final punctuation, in order to avoid confusing it with
%%% the Web address.
%%%
%%% To suppress output of a particular field, define its macro to expand
%%% to an empty string, or better, \unskip, like this:
%%%
%%% \newcommand{\showDOI}[1]{\unskip}   % LaTeX syntax
%%%
%%% \def \showDOI #1{\unskip}           % plain TeX syntax
%%%
%%% ====================================================================

\ifx \showCODEN    \undefined \def \showCODEN     #1{\unskip}     \fi
\ifx \showDOI      \undefined \def \showDOI       #1{#1}\fi
\ifx \showISBNx    \undefined \def \showISBNx     #1{\unskip}     \fi
\ifx \showISBNxiii \undefined \def \showISBNxiii  #1{\unskip}     \fi
\ifx \showISSN     \undefined \def \showISSN      #1{\unskip}     \fi
\ifx \showLCCN     \undefined \def \showLCCN      #1{\unskip}     \fi
\ifx \shownote     \undefined \def \shownote      #1{#1}          \fi
\ifx \showarticletitle \undefined \def \showarticletitle #1{#1}   \fi
\ifx \showURL      \undefined \def \showURL       {\relax}        \fi
% The following commands are used for tagged output and should be
% invisible to TeX
\providecommand\bibfield[2]{#2}
\providecommand\bibinfo[2]{#2}
\providecommand\natexlab[1]{#1}
\providecommand\showeprint[2][]{arXiv:#2}

\bibitem[\protect\citeauthoryear{Axelsen and Yokoyama}{Axelsen and
  Yokoyama}{2015}]%
        {AxYo15}
\bibfield{author}{\bibinfo{person}{Axelsen, H.B.} {and}
  \bibinfo{person}{Yokoyama, T}.} \bibinfo{year}{2015}\natexlab{}.
\newblock \showarticletitle{Programming Techniques for Reversible Comparison
  Sorts}. In \bibinfo{booktitle}{\emph{APLAS}} \emph{(\bibinfo{series}{LNCS})},
  \bibfield{editor}{\bibinfo{person}{Feng, X.} {and} \bibinfo{person}{Park, S.}} (Eds.), Vol.~\bibinfo{volume}{9458}.
  \bibinfo{publisher}{Springer-Verlag}, \bibinfo{pages}{407--426}.
\newblock


\bibitem[\protect\citeauthoryear{Buhrman, Tromp, and Vit{\'a}nyi}{Buhrman
  et~al\mbox{.}}{2001}]%
        {BuTV01}
\bibfield{author}{\bibinfo{person}{Buhrman, H.}, \bibinfo{person}{Tromp, J.}, {and} \bibinfo{person}{Vit{\'a}nyi}, P.}
  \bibinfo{year}{2001}\natexlab{}.
\newblock \showarticletitle{Time and Space Bounds for Reversible Simulation}.
  In \bibinfo{booktitle}{\emph{ICALP}} \emph{(\bibinfo{series}{LNCS})},
  \bibfield{editor}{\bibinfo{person}{Orejas, F.}, \bibinfo{person}{Spirakis, P.G.}, {and} \bibinfo{person}{van Leeuwen, J.}} (Eds.),
  Vol.~\bibinfo{volume}{2076}. \bibinfo{publisher}{Springer-Verlag},
  \bibinfo{pages}{1017--1027}.
\newblock


\bibitem[\protect\citeauthoryear{Frank}{Frank}{1999}]%
        {Fran99}
\bibfield{author}{\bibinfo{person}{Frank, M.P}.}
  \bibinfo{year}{1999}\natexlab{}.
\newblock \emph{\bibinfo{title}{Reversibility for Efficient Computing}}.
\newblock \bibinfo{thesistype}{Ph.D. Dissertation}. \bibinfo{school}{MIT}.
\newblock


\bibitem[\protect\citeauthoryear{Haulund, Mogensen, and Gl{\"u}ck}{Haulund
  et~al\mbox{.}}{2017}]%
        {HaMG17}
\bibfield{author}{\bibinfo{person}{Haulund, T.},
  \bibinfo{person}{Mogensen, T.{\AE}.}, {and} \bibinfo{person}{Gl{\"u}ck}, R.} \bibinfo{year}{2017}\natexlab{}.
\newblock \showarticletitle{Implementing Reversible Object-Oriented Language
  Features on Reversible Machines}. In \bibinfo{booktitle}{\emph{RC}},
  \bibfield{editor}{\bibinfo{person}{Phillips, I.} {and}
  \bibinfo{person}{Rahaman, H.}} (Eds.).
  \bibinfo{publisher}{Springer-Verlag}, \bibinfo{pages}{66--73}.
\newblock
\showISBNx{978-3-319-59936-6}


\bibitem[\protect\citeauthoryear{Morita}{Morita}{2017}]%
        {Mori17a}
\bibfield{author}{\bibinfo{person}{Morita, K}.}
  \bibinfo{year}{2017}\natexlab{}.
\newblock \bibinfo{booktitle}{\emph{Theory of Reversible Computing}}.
\newblock \bibinfo{publisher}{Springer-Verlag}.
\newblock


\bibitem[\protect\citeauthoryear{Yokoyama, Axelsen, and Gl\"{u}ck}{Yokoyama
  et~al\mbox{.}}{2008}]%
        {YoAG08a}
\bibfield{author}{\bibinfo{person}{Yokoyama, T},
  \bibinfo{person}{Axelsen, H.B.}, {and} \bibinfo{person}{Gl\"{u}ck}, R.} \bibinfo{year}{2008}\natexlab{}.
\newblock \showarticletitle{Principles of a Reversible Programming Language}.
  In \bibinfo{booktitle}{\emph{Computing Frontiers. Proceedings}}.
  \bibinfo{publisher}{ACM {Press}}, \bibinfo{pages}{43--54}.
\newblock


\end{thebibliography}
%abbrev,%
%ref,%
%ref-jp,%
%yokoyama-jp,%
%yokoyama%
%}

%\appendix
%\section{Appendix}
%\begin{sloppypar}
%All the programs presented in this paper can be executed on our online interpreter available at \url{http://tetsuo.jp/janus-playground/}.
%\end{sloppypar}

%\begin{figure}[h]
%\centering
%\includegraphics[width=.40\textwidth]{img/input-output.eps}
%\caption{The input and output of reversible search algorithms to consider.}\label{fig:inout}
%\end{figure}

%\begin{figure*}[h]
%\includegraphics[bb=0 670 1680 783,width=.99\textwidth]{img/pointers_yoko.eps}
%\caption{Reversible traversal of a doubly linked list.}\label{fig:pointers}
%\end{figure*}

\end{document}